\documentclass[article,preprint,groupedaddress]{revtex4}
\usepackage{epsfig}
\usepackage{graphicx}
\usepackage{amssymb}

\begin{document}

\title{Extremal black holes have external light rings}
\author{Shahar Hod}
\address{The Ruppin Academic Center, Emeq Hefer 40250, Israel}
\address{}
\address{The Hadassah Institute, Jerusalem 91010, Israel}
\date{\today}

\begin{abstract}
\ \ \ It is proved that spherically symmetric extermal black holes 
possess at least one external light ring. Our remarkably compact proof is 
based on the dominant energy condition which characterizes the external matter fields in the non-vacuum extremal 
black-hole spacetimes.  
\end{abstract}
\bigskip
\maketitle

\section{Introduction}

The non-linearly coupled Einstein-matter field equations predict, under plausible 
physical assumptions \cite{Bar,Chan,Shap,Hodub,Herne}, that curved spacetimes of highly compact objects are characterized by the presence of closed null 
circular trajectories (external light rings). The fact that massless particles can perform 
orbital motions along closed circular geodesics plays a key role in understanding many of the fundamental physical properties 
of the corresponding central compact objects \cite{Bar,Chan,Shap,Herne,Hodns,Mash,Goeb,Hod1,Dec,Hodhair,Hodfast,YP,Hodm,Hodub,Lu1,Hodlwd,Pod,Ame,Ste}. 

In particular, the intriguing physical phenomenon of strong gravitational lensing, 
which can be used as an important observational tool to identify the existence of cosmological black holes, 
is a direct outcome of the presence of closed null geodesics in the highly curved near-horizon 
regions of the corresponding black-hole spacetimes \cite{Pod,Ame,Ste}. 
In addition, it is well established (see \cite{Mash,Goeb,Hod1,Dec} and references therein) 
that the relaxation rates of perturbed black-hole spacetimes are closely related 
to the instability timescales that characterize the geodesic motions of massless particles along the 
null circular geodesics of the corresponding curved spacetimes. 

Interestingly, it has been proved \cite{Hodfast,YP} that, as judged by far away asymptotic observers, 
the innermost null circular geodesic of 
a black-hole spacetime provides the fastest way to travel around the central black hole. 
In addition, it has been revealed \cite{Hodhair,Hodub,Hodlwd,Hod1} that, 
in spherically symmetric hairy (non-vacuum) black-hole spacetimes, the effective 
lengths of the external matter fields are bounded from below by the radii of the 
innermost null circular geodesics that characterize the corresponding curved spacetimes. 

The fact that null circular geodesics have a significant role 
in determining many of the fundamental physical properties of black-hole spacetimes 
naturally raises the following important question: Do the Einstein-matter field equations guarantee the existence of 
external null circular trajectories (light rings) in all black-hole spacetimes? 
Intriguingly, the existence of closed null circular geodesics in the external regions of 
asymptotically flat {\it non}-extremal spherically symmetric black-hole spacetimes 
has been proved in \cite{Hodub}. A general (and mathematically elegant) proof 
for the existence of null circular geodesics in {\it non}-extremal 
stationary axi-symmetric black-hole spacetimes has recently 
been presented in the physically interesting work \cite{Herne}. 

It is important to point out that the theorems presented in \cite{Hodub,Herne} for the existence of external 
null circular geodesics in generic black-hole spacetimes seem to fail for {\it extremal} hairy (non-vacumm) black-hole 
configurations. 
Motivated by this fact, it has recently been proved \cite{Hoddo} 
that extermal black-hole spacetimes with positive tangential pressures on their horizons [$p_{\text{T}}(r=r_{\text{H}})>0$, see Eq. (\ref{Eq2}) below] possess external light rings. 
However, as emphasized in \cite{Hoddo}, 
the existence theorem presented in \cite{Hoddo} is not valid for extermal black holes with non-positive 
horizon tangential pressures. 
This fact leaves open the possibility of finding extremal black-hole spacetimes with non-positive horizon tangential 
pressures that do not have external light rings.

The main goal of the present compact paper is to complete our knowledge about the (in)existence 
of external null circular geodesics in extremal black-hole spacetimes. 
In particular, using analytical techniques, we shall explicitly prove below that 
the non-linearly coupled Einstein-matter field equations guarantee 
that spherically symmetric hairy (non-vacuum) extermal black-hole spacetimes whose external matter fields 
respect the dominant energy condition are always characterized by the presence of 
external null circular geodesics (closed light rings). 

\section{Description of the system}

We consider spherically symmetric extremal black-hole spacetimes which, using 
the familiar Schwarzschild spacetime coordinates $\{t,r,\theta,\phi\}$, are 
characterized by the curved line element
\cite{Hodfast,Hodm,Noteunit}
\begin{equation}\label{Eq1}
ds^2=-e^{-2\delta}\mu dt^2 +\mu^{-1}dr^2+r^2(d\theta^2 +\sin^2\theta d\phi^2)\  ,
\end{equation}
where the radially-dependent functions $\mu=\mu(r)$ and $\delta=\delta(r)$ are determined by 
the matter content of the non-vacuum spacetime. 

Using the line element (\ref{Eq1}) and the notations \cite{Bond1}
\begin{equation}\label{Eq2}
\rho\equiv-T^{t}_{t}\ \ \ \ ,\ \ \ \ p\equiv T^{r}_{r}\ \ \ \ , \ \ \ \ p_T\equiv T^{\theta}_{\theta}=T^{\phi}_{\phi}\
\end{equation}
for the radially-dependent energy density $\rho$, radial pressure $p$, and tangential pressure $p_T$ 
of the external static matter configurations, one finds that the Einstein-matter field 
equations $G^{\mu}_{\nu}=8\pi T^{\mu}_{\nu}$ yield the radial differential equations \cite{Hodfast,Hodm}
\begin{equation}\label{Eq3}
{{d\mu}\over{dr}}=-8\pi r\rho+{{1-\mu}\over{r}}\
\end{equation}
and
\begin{equation}\label{Eq4}
{{d\delta}\over{dr}}=-{{4\pi r(\rho +p)}\over{\mu}}\  ,
\end{equation}
which relate the metric functions to the external matter sources. 

Extremal black-hole spacetimes are characterized by the horizon boundary conditions \cite{Bekreg}:
\begin{equation}\label{Eq5}
\mu(r=r_{\text{H}})=0\  ,
\end{equation}
\begin{equation}\label{Eq6}
\Big[{{d\mu}\over{dr}}\Big]_{r=r_{\text{H}}}=0\  ,
\end{equation}
\begin{equation}\label{Eq7}
\Big[{{d^2\mu}\over{dr^2}}\Big]_{r=r_{\text{H}}}>0\  ,
\end{equation}
\begin{equation}\label{Eq8}
\delta(r=r_{\text{H}})<\infty\ \ \ ; \ \ \ \Big[{{d\delta}\over{dr}}\Big]_{r=r_{\text{H}}}<\infty\  ,
\end{equation}
and
\begin{equation}\label{Eq9}
p(r=r_{\text{H}})=-\rho(r=r_{\text{H}})=-(8\pi r^2_{\text{H}})^{-1}\  .
\end{equation}
In addition, the radially-dependent metric functions of 
asymptotically flat black-hole spacetimes are characterized by the functional relations 
\begin{equation}\label{Eq10}
\mu(r\to\infty)\to1\ 
\end{equation}
and
\begin{equation}\label{Eq11}
\delta(r\to\infty)\to0\  .
\end{equation}

Taking cognizance of the Einstein equation (\ref{Eq3}), one finds the expression
\begin{equation}\label{Eq12}
\mu(r)=1-{{2m(r)}\over{r}}\ 
\end{equation}
for the dimensionless metric function $\mu(r)$, where
\begin{equation}\label{Eq13}
m(r)={{r_{\text{H}}}\over{2}}+\int_{r_{\text{H}}}^{r} 4\pi r^{2}\rho(r)dr\
\end{equation}
is the gravitational mass contained within a sphere of radius $r$ [here 
$m(r=r_{\text{H}})=r_{\text{H}}/2$ is the 
mass contained within the black-hole horizon]. 
Taking cognizance of Eqs. (\ref{Eq10}), (\ref{Eq12}), and (\ref{Eq13}), one deduces the 
characteristic functional relation 
\begin{equation}\label{Eq14}
r^3\rho(r)\to 0\ \ \ \ \text{for}\ \ \ \ r\to\infty\
\end{equation}
for the external energy density in asymptotically flat black-hole spacetimes.

Our proof, to be presented below, for the existence of external null circular geodesics in extremal black-hole spacetimes 
is based on the well known dominant energy condition which, for a given density profile of the matter fields, 
bounds from above the (radial and tangential) pressure components of the corresponding matter distribution \cite{Bekreg}:
\begin{equation}\label{Eq15}
|p|, |p_T|\leq\rho\  .
\end{equation}

\section{The proof for the existence of external null circular geodesics in extremal black-hole spacetimes}

In the present section we shall explicitly prove that extremal hairy (non-vacuum) 
black-hole spacetimes that respect the dominant energy condition (\ref{Eq15}) 
necessarily possess at least one external light ring whose radius satisfies the inequality 
$r_{\gamma}>r_{\text{H}}$. 

To this end, we shall analyze the radial functional behavior of the dimensionless function 
\begin{equation}\label{Eq16}
{\cal N}(r)\equiv 3\mu-1-8\pi r^2p\
\end{equation}
in the extremal black-hole spacetime (\ref{Eq1}). It has been explicitly proved \cite{Hodhair} 
that, in spherically symmetric black-hole spacetimes, the radii of null circular geodesics are determined 
by the mathematically compact relation  
\begin{equation}\label{Eq17}
{\cal N}(r=r_{\gamma})=0\  .
\end{equation}

Taking cognizance of Eqs. (\ref{Eq5}), (\ref{Eq9}), and (\ref{Eq16}), one deduces the boundary relation 
\begin{equation}\label{Eq18}
{\cal N}(r=r_{\text{H}})=0\
\end{equation}
on the outer horizon of the extremal black hole. 
In addition, the functional relation (\ref{Eq14}), which characterizes asymptotically flat 
black-hole spacetimes, together with the assumed dominant energy condition (\ref{Eq15}), imply 
the asymptotic functional behavior
\begin{equation}\label{Eq19}
r^3p(r)\to 0\ \ \ \ \text{for}\ \ \ \ r\to\infty\
\end{equation}
for the external radial pressure. 
From Eqs. (\ref{Eq10}) and (\ref{Eq19}) one finds the characteristic large-$r$ behavior 
\begin{equation}\label{Eq20}
{\cal N}(r\to\infty)\to 2\
\end{equation}
of the dimensionless radial function (\ref{Eq16}). 

Defining the dimensionless function 
\begin{equation}\label{Eq21}
{\cal F}\equiv r{{d\mu}\over{dr}}\
\end{equation}
and using Eqs. (\ref{Eq6}) and (\ref{Eq7}), one finds the characteristic relation
\begin{equation}\label{Eq22}
\Big[{{d{\cal F}}\over{dr}}\Big]_{r=r_{\text{H}}}=r_{\text{H}}\Big[{{d^2\mu}\over{dr^2}}\Big]_{r=r_{\text{H}}}>0\
\end{equation}
for extremal black holes. 
In addition, from the Einstein equation (\ref{Eq3}) and the boundary condition (\ref{Eq6}) 
one obtains the horizon relation 
\begin{equation}\label{Eq23}
\Big[{{d{\cal F}}\over{dr}}\Big]_{r=r_{\text{H}}}=-{{d}\over{dr}}[8\pi r^2\rho]_{r=r_{\text{H}}}\
\end{equation}
for the extremal black-hole spacetime (\ref{Eq1}). 
From Eqs. (\ref{Eq22}) and (\ref{Eq23}) one deduces that the dimensionless function $r^2\rho$ 
decreases in the vicinity of the extremal black-hole horizon:
\begin{equation}\label{Eq24}
\Big[{{d{(r^2\rho)}}\over{dr}}\Big]_{r=r_{\text{H}}}<0\  .
\end{equation}

Taking cognizance of the horizon boundary relation (\ref{Eq9}) and the assumed 
dominant energy condition (\ref{Eq15}), one deduces from (\ref{Eq24}) that the radial expression $r^2p$ is 
a negative increasing function in the vicinity of the black-hole horizon:
\begin{equation}\label{Eq25}
[r^2p]_{r=r_{\text{H}}}<0\ \ \ \ \text{and}\ \ \ \ \Big[{{d{(r^2p)}}\over{dr}}\Big]_{r=r_{\text{H}}}>0\  .
\end{equation}

From the analytically derived functional relation (\ref{Eq25}) and the horizon boundary condition (\ref{Eq6}) 
for extremal black holes, one finds the characteristic inequality [see Eq. (\ref{Eq16})]
\begin{equation}\label{Eq26}
\Big[{{d{\cal N}}\over{dr}}\Big]_{r=r_{\text{H}}}=-8\pi\Big[{{d{(r^2p)}}\over{dr}}\Big]_{r=r_{\text{H}}}<0\  ,
\end{equation}
which, together with the relation (\ref{Eq18}), imply that the dimensionless radial function (\ref{Eq16}) is {\it non}-positive 
in the vicinity of the black-hole horizon. 
In particular, the radial function ${\cal N}(r)$ is characterized by the near-horizon property:
\begin{equation}\label{Eq27}
{\cal N}(r/r_{\text{H}}\to 1^+)\to 0^{-}\  .
\end{equation}

Finally taking cognizance of the analytically derived near-horizon relation (\ref{Eq27}) 
and the characteristic asymptotic behavior (\ref{Eq20}) of the dimensionless radial function (\ref{Eq16}), 
one deduces that spherically symmetric extremal black-hole spacetimes 
whose external matter fields respect the dominant energy condition (\ref{Eq15}) 
possess at least one external null circular geodesic (closed light ring) which 
is characterized by the functional relation
\begin{equation}\label{Eq28}
{\cal N}(r=r_{\gamma})=0\ \ \ \ \text{with}\ \ \ \ r_{\gamma}>r_{\text{H}}\  .
\end{equation}

\section{Summary}

Null circular geodesics play important roles in fundamental as well as observational studies of the 
physics of curved black-hole spacetimes (see \cite{Bar,Chan,Shap,Herne,Hodns,Mash,Goeb,Hod1,Dec,Hodhair,Hodfast,YP,Hodub,Lu1,Hodlwd,Pod,Ame,Ste} and references 
therein). 
Interestingly, the existence of closed light rings in asymptotically flat {\it non}-extremal 
black-hole spacetimes has been proved, using the Einstein-matter field equations, 
in \cite{Hodub} for spherically symmetric non-vacuum (hairy) black-hole configurations. 
A mathematically elegant proof for the existence of external null circular geodesics in {\it non}-extremal 
stationary axi-symmetric black-hole spacetimes has been provided in the highly important work \cite{Herne}. 

Intriguingly, the existence theorems presented in \cite{Hodub,Herne} seem to fail for {\it extremal} 
black-hole spacetimes. Motivated by this observation, we have raised 
the following physically important question \cite{Hoddo}: Do extremal black-hole spacetimes always possess external 
light rings? 

In the present paper we have presented a remarkably compact theorem \cite{Noteafc}, which is based on 
the non-linearly coupled Einstein-matter field equations, that reveals the physically important fact that 
spherically symmetric extermal hairy (non-vacuum) 
black holes whose external matter fields respect the dominant energy condition necessarily 
possess at least one external light ring (closed null circular geodesic). 


\bigskip
\noindent {\bf ACKNOWLEDGMENTS}

This research is supported by the Carmel Science Foundation. I thank
Yael Oren, Arbel M. Ongo, Ayelet B. Lata, and Alona B. Tea for
stimulating discussions.


\begin{thebibliography}{99}
\bibitem{Noteunit} We use natural units in which $G=c=1$.
\end{thebibliography}
\end{document}